\documentclass[twocolumn,10pt]{IEEEtran}

\usepackage{amsmath,graphicx}

\usepackage{multirow}
\usepackage{longtable}
\usepackage{amsthm}
\usepackage{amssymb }
\usepackage{amsmath}
\usepackage{bm}
\usepackage{color}
\usepackage{mathrsfs}
\usepackage{amsfonts}
\usepackage{longtable}
\usepackage{multirow}
\usepackage{algorithm, algpseudocode}
\usepackage{cite}
\usepackage{url}
\usepackage{tabularx}
\usepackage{lscape}
\usepackage{forest}
\usepackage{array}
\usepackage{xcoffins}
\usepackage{subfigure}
\usepackage{flushend}

\NewCoffin\tablecoffin
\NewDocumentCommand\Vcentre{m}
  {%
    \SetHorizontalCoffin\tablecoffin{#1}%
    \TypesetCoffin\tablecoffin[l,vc]%
  }

\usetikzlibrary{shadows}
\newcolumntype{C}[1]{>{\centering\arraybackslash}p{#1}}

\newcolumntype{b}{>{\hsize=2.3\hsize}X}
\newcolumntype{s}{>{\hsize=.45\hsize}X}
\newcolumntype{m}{>{\hsize=.9\hsize}X}

\def\LOTUS{\text{LOTUS}}
\def\VIP{\text{VIP-CUP}}

\def\Y{\bm{Y}}
\def\X{\bm{X}}

\title{Lung-Originated Tumor Segmentation from Computed Tomography Scan (LOTUS) Benchmark\thanks{Corresponding Author is Arash Mohammadi, email: arash.mohammadi@concordia.ca. Due to the space limitation, affiliations appear in the acknowledgment section. Parnian Afshar, Arash Mohammadi, Konstantinos N. Plataniotis, Keyvan Farahani,  Justin Kirby, Anastasia Oikonomou, Amir Asif, Leonard Wee, Andre Dekker,  and  Xin Wu are the co-organizers of 2018 IEEE VIP-Cup competition, and others have participated in the competition.}}
\author{Parnian Afshar, Arash Mohammadi, Konstantinos N. Plataniotis, Keyvan Farahani,  Justin Kirby, Anastasia Oikonomou, Amir Asif, Leonard Wee, Andre Dekker, Xin Wu , Mohammad Ariful Haque, Shahruk Hossain, Md. Kamrul Hasan, Uday Kamal, Winston Hsu, Jhih-Yuan Lin, M. Sohel Rahman, Nabil Ibtehaz, Sh. M. Amir Foisol, Kin-Man Lam, Zhong Guang, Runze Zhang, Sumohana S. Channappayya, Shashank Gupta and Chander Dev\\}

\begin{document}
\date{\today}
\maketitle
\begin{abstract}
Lung cancer is one of the deadliest cancers, and in part its effective diagnosis and treatment depend on the accurate delineation of the tumor. Human-centered segmentation, which is currently the most common approach, is subject to inter-observer variability, and is also time-consuming, considering the fact that only experts are capable of providing annotations. Automatic and semi-automatic tumor segmentation methods have recently shown promising results. However, as different researchers have validated their algorithms using various datasets and performance metrics, reliably evaluating these methods is still an open challenge. The goal of the Lung-Originated Tumor Segmentation from Computed Tomography Scan (LOTUS) Benchmark created through 2018 IEEE Video and Image Processing (VIP) Cup competition, is to provide a unique dataset and pre-defined metrics, so that different researchers can develop and evaluate their methods in a unified fashion. The 2018 VIP Cup started with a global engagement from 42 countries to access the competition data. At the registration stage, there were 129 members clustered into 28 teams from 10 countries, out of which 9 teams made it to the final stage and 6 teams successfully completed all the required tasks. In a nutshell, all the algorithms proposed during the competition, are based on deep learning models combined with a false positive reduction technique. Methods developed by the three finalists show promising results in tumor segmentation, however, more effort should be put into reducing the false positive rate. This competition manuscript presents an overview of the VIP-Cup challenge, along with the proposed algorithms and results.
\end{abstract}
\section{Introduction} \label{sec:Introduction}
Lung cancer is among the most common cancers, responsible for most of the cancer-related fatalities, all over the world~\cite{Gridelli:2015}. About $85\%$ of the diagnosed lung cancer cases are categorized as Non-small-cell lung cancer (NSCLC), which is mainly caused by smoking, radon exposure, and air pollution. Although this type of cancer is widespread, it is, most of the times, diagnosed in advanced stages, where it is too late for pursuing an effective and optimal treatment. The late diagnosis of the NSCLC, which is mainly due to the lack of earlier clinical symptoms, negatively affects the quality of patients' lives. This calls for an urgent quest to further develop sensitive screening  technologies together with advanced processing solutions to provide accurate diagnosis in early stages.

According to the American Cancer Society~\cite{American:2016}, Computed Tomography (CT) scan has the potential to reveal lung tumors and their associated position and shape, which are the essential requirements for providing accurate diagnosis in early stages. Therefore, people with high risks of developing lung tumors should undergo adequate CT screening, to increase the chance of detecting the lung tumor in early stages, where there are no clinical symptoms to help with the diagnosis.
The accurate delineation (segmentation) of the lung tumor, from the CT images, is the first step in analyzing the tumor and making decisions on the type of the required treatment~\cite{Gu:2013}. The process of segmenting the nodule (referred to as tumor, if it is cancerous) from the medical images, extracting quantitative and semi-quantitative features from the segmented regions, and analyzing them with the ultimate goal of diagnosis/prediction is referred to as ``radiomics''~\cite{Aerts:2014, Parnian:SPM18}.
In other words, radiomics features and landmarks that are needed to reveal the nature of the tumor are mainly extracted from the annotated abnormality.
 The quality of the extracted features, therefore, highly depends on the accuracy of the segmented region. Furthermore, several treatments, such as radiotherapy, require the exact position and boundaries of the tumor~\cite{Gu:2013}. Despite the crucial need for the accurate tumor segmentation, manual annotation (involving experts and radiologists) is the most commonly used approach, which is highly subject to inter-observer variability~\cite{Griethuysen:2017}, reducing the stability and reliability of the extracted radiomics features. Besides, to make sure that the entire tumor volume will be removed during the selected treatment, experts, typically, tend to overestimate the tumor boundaries~\cite{Gu:2013} resulting in removal of healthy tissues. This further necessitates the need for more robust, reliable, and accurate segmentation techniques.

\begin{table*}[t!]
\caption{\small Different methods applied to lung tumor/nodule segmentation along with the dataset and performance metric.}
\label{table:lit}
\centering
\begin{tabular}{ |c|c|c|c| }
\hline
\textbf{Author}& \textbf{Dataset (Number of the subjects)} & \textbf{Method} & \textbf{Performance Metric}  \\
\hline
Gu \em{et. al.}\normalfont  \cite{Gu:2013}  & In-house dataset ($129$) & Region growing & Dice score\\
\hline
Dehmeshki \em{et. al.}\normalfont  \cite{Dehmeshki:2008} &In-house dataset ($423$)& Region growing& Validated by the radiologists\\
\hline
Kubota \em{et. al.} \normalfont \cite{Kubota:2011} & LIDC-IDRI ($1010$)~\cite{Armato:2011,Armato:2015,Clark:2013} & Region growing & Dice score\\
\hline
Diciotti \em{et. al.}\normalfont  \cite{Diciotti:2011}&  In-house dataset ($202$) \& LIDC-IDRI & Morphological operation & Validated by the radiologists\\
\hline
Messay \em{et. al.} \normalfont \cite{Messay:2010}& In-house dataset ($90$) \& LIDC-IDRI&  Morphological operation & Dice score\\
\hline
Wang \em{et. al.} \normalfont \cite{Wang:2017}& In-house dataset ($74$) \& LIDC-IDRI & CNN & Dice score \& average boundary distance\\
\hline
Farag \em{et. al.} \normalfont \cite{Farag:2013}  & ELCAP ($50$)~\cite{Reeves:2017} \& LIDC-IDRI & Level-set & ROC\\
\hline
Ye \em{et. al.} \normalfont \cite{Ye:2010}  & In-house dataset ($181$) & Graph-cut & Dice score\\
\hline
Keshani \em{et. al.} \normalfont \cite{Keshani:2013}  & ANODE09 ($55$)\cite{Ginneken:2010} \& LIDC-IDRI& Active contour & Dice score\\
\hline
Song \em{et. al.} \normalfont \cite{Song:2016}  & In-house dataset ($100$) \& LIDC-IDRI& Region growing & Dice score \& Hausdorff distance\\
\hline
Afshar \em{et. al.} \normalfont \cite{Afshar:2016}  & LIDC-IDRI & Clustering & Entropy analysis\\
\hline
Wu \em{et. al.} \normalfont \cite{Wu:2010}  & In-house dataset ($239$) & CRF & Prediction accuracy\\
\hline
Mukherjee \em{et. al.} \normalfont \cite{Mukherjee:2017}  & LIDC-IDRI & Graph-cut \& Deep learning & Dice score\\
\hline
Ciompi \em{et. al.} \normalfont \cite{Ciompi:2017}   & MILD ($943$)~\cite{Pastorino:2012} \& DLCST ($468$)~\cite{Pedersen:2009} & CNN & accuracy \& F-measure\\
\hline
Setio \em{et. al.} \normalfont \cite{Setio:2017} & ANODE09 \& LIDC-IDRI \& DLCST & CNN & ROC \& Competition Performance Metric\\
\hline
Feng \em{et. al.} \normalfont \cite{Feng:2017} &  LIDC-IDRI & CNN & Dice score\\
\hline
\end{tabular}
\end{table*}
Automatic tumor segmentation has the promise of generating more robust and accurate segmentation results taking the burden of manual delineation of the experts. However, there are several challenges in the way of having accurate and real-time automatic tumor segmentation. First of all, there are, typically, limited visible landmarks and intensity differences between normal and abnormal tissues~\cite{Menze:2015} making, even a manual annotation, difficult and error-prone. Furthermore, shape and spatial priors (commonly used  for image segmentation in other multi-media domains) are not applicable for  tumor segmentation, because there are high varieties of tumor shapes and types among different patients~\cite{Menze:2015}. More importantly, in spite of the efforts to create standard platforms and protocols across  different imaging institutions, there are still several incompatibilities that make development of unique segmentation solutions a challenging task.
Considering the advantages and, at the same time, the challenges of the automatic lung tumor segmentation, there have been ongoing studies and investigations~\cite{Ju:2015} on development of accurate tumor segmentation methods. Although these methods have shown promising results on specific and small (er) datasets, the fact that outcomes are reported based on different metrics and datasets has made different approaches incomparable, thereby limiting their clinical application.

The goal of the $\LOTUS$ benchmark is to provide an opportunity to develop automatic lung tumor segmentation methods based on a unique dataset. The $\LOTUS$ benchmark was constructed through the 2018 Video \& Image Processing (VIP) Cup competition, which was organized in conjunction with 2018 IEEE International Conference on Image Processing (ICIP) to test and evaluate different algorithms against pre-defined measures. In this paper, we describe the $\LOTUS$ benchmark setup and results, along with the details of the developed methods by the final six participating teams.


\section{Prior works on Lung Tumor Segmentation} \label{sec:state}
Table~\ref{table:lit} presents several works on the problem of lung tumor segmentation from different perspectives. As it can be inferred from Table~\ref{table:lit}, researchers have used various datasets and performance metrics, which make the comparison of their results almost impossible. This calls for a venue, where all algorithms are tested on a unique dataset and based on unique evaluation criteria, which is the target of this manuscript. Before introducing the $\LOTUS$ setup, first different segmentation methods specified in Table~\ref{table:lit} are briefly described.

In general, studies on lung tumor segmentation can be categorized into two groups of conventional and deep learning-based methods. Conventional or traditional approaches are developed by considering a set of pre-defined features capable of classifying pixels as being part of a tumor or not. Deep learning-based methods, on the other hand, do not require extra information on what types of features to consider for distinguishing the image pixels. In other words, these methods can attempt to learn the best features, on their own, with the goal of maximizing the segmentation accuracy. The question raised naturally at this point is that what would be the criteria to demonstrate that maximum accuracy is achieved (this would be a question in both conventional and deep learning-based methods). Intuitively speaking, the minimum error (resulting in the maximum accuracy) is zero, i.e., the segmented region and the ground truth completely match. However, in reality this scenario can not occur, because of the facts that training is performed on a sub-sample of all the possible inputs (decreasing the chance of capturing the correct distribution), presence of noise, and limitation of the time and computational resources.
Therefore, the aforementioned question makes sense and an alternative mechanism is required. One solution, would be to find quantitative lower bounds  on the error (either analytically or empirically) representing the maximum achievable accuracy.

Below, we briefly describe the conventional and deep learning-based approaches along with the most common techniques that fall within these two categories. It is worth mentioning that the tumor segmentation is sometimes followed by a false positive (FP) reduction phase~\cite{Keshani:2013}, where several features are extracted from the segmented regions and, by applying a classifier such as a Support Vector Machine (SVM), non-nodule regions are excluded. Such an approach is also adopted in~\cite{Lu:2011}, where different features, including size, diameter, and boundary smoothness, are extracted, in order to detect positive lesion candidates.

\subsection{Conventional Lung Tumor Segmentation Methods} \label{sec:conv}
Conventional segmentation approaches used for the problem of lung tumor segmentation can be divided into three main categories: Intensity-based, Model-based and Machine learning approaches.
Below we briefly explain these three categories with special focus on their application for lung tumor segmentation tasks.

\vspace{.025in}
\noindent
\textbf{\textit{Intensity-based Segmentation:}}
Methods belonging to this category use the intensity of the pixels as the indicator of whether they belong to the tumor region or not. In other words, pixels are classified simply based on their intensity values and the differences between the intensity values of the neighboring pixels. The threshold, according to which pixels are classified, is typically derived from the intensity histogram~\cite{Natarajan:2012}. Region growing approach is one of the most popular intensity-based methods in tumor segmentation~\cite{Dehmeshki:2008, Kubota:2011, Song:2016}. This approach basically consists of first selecting one or more seeds within the tumor region, followed by computing the connectivity between the seeds and their associated neighbors. This connectivity, which is constructed based on the intensity differences of the pixels, is then used to extend the tumor region.
There are, however, several challenges in lung tumor segmentation~\cite{Dehmeshki:2008} that limits the applicability of the classical region growing approach. First of all, some types of the lung tumors are attached to blood vessels with very similar intensities. Second, the tumor may have a low contrast with its background, and finally, classical region growing methods are highly sensitive to the initial seeds, causing difficulty in providing reproducible outcomes.

Due to the aforementioned drawbacks of the region growing techniques, and intensity-based methods in general, typically, different morphological operations~\cite{Diciotti:2011} are applied to the initial segmentation, with the goal of detaching the tumor from the wrongly selected vessels and background. These operations can be designed by considering different shape priors for the tumor and other components. For instance, tumors tend to be spheroidal objects, whereas blood vessels have long structures. Such a morphological operation is investigated in~\cite{Messay:2010}, where a circular structuring element is used to remove vessels and other attachments. Furthermore, to reduce the false positive rate (FPR) in detecting tumor candidates, a rule-based analysis is applied based on extracting features, such as area, volume and diameter. Nevertheless, irregular shaped nodules and various tumor sizes limit the generalization capability of the morphological operations~\cite{Wang:2017}.

\vspace{.025in}
\noindent
\textbf{\textit{Model-based Segmentation:}}
Model-based~\cite{Lassen:2015} (also referred to as Energy-based) techniques start with defining a shape prior over the region of interest (ROI). By defining and optimizing an energy function, composed of different terms such as image contrast and image gradient forces, the prior can evolve and fit the ROI. One of the major limitations of the model-based approaches is that the formulation may require large number of parameters, that are difficult to tune~\cite{Farag:2013}.

Active contours~\cite{Xu:2007} (also referred to as Snakes) are widely used within model-based segmentation methods, that start by explicitly defining several control points/particles on an initial contour. The contour evolves to fit the boundary of the ROI, with the aim of minimizing an energy function, consisting of an internal and an external energy term. The internal term aims to maintain a smooth contour, while the external term tries to pull the contour towards optimal image features. Active contours have a major drawback that limits their applicability for lung tumor segmentation, i.e., they cannot support the splitting or merging of the contours, leading to their failure when the tumor does not have an integrated pattern (joined shape).
To overcome this shortcoming, level-set segmentation~\cite{Farag:2013} is proposed, which defines the contour, implicitly, as the zero level-set of a surface. Evolving the surface instead of the explicit contour points has the advantage of handling the situations, where contours need to split or merge to fit the correct tumor boundary.

Graph cut method~\cite{Ye:2010} is another technique within the category of energy-based models, which has been widely investigated to solve the problem of lung tumor segmentation. Graph cut approach is based on constructing a graph, whose nodes correspond to image pixels, and edge weights correspond to features such as the similarity of neighboring pixels. The problem is, then, defined as finding a minimum cut in the graph that goes through the ROI boundary. Similar to intensity-based approaches, the performance of the energy-based models is limited when dealing with low contrast tumors or tumors attached to the vessels~\cite{Wang:2017}.

\vspace{.025in}
\noindent
\textbf{\textit{Machine Learning-based Segmentation:}}
Machine learning techniques for lung tumor segmentation consist of extracting a set of pre-designed features from the pixels/voxels or patches (groups of neighboring pixels), and learning the labels, i.e., being a part of the nodule or not, in a supervised or unsupervised fashion. In the former case, labels are learned from a set of training data, with available ground truth, whereas in the latter case, labels are assigned to pixels, based on their similarity (i.e., the ground truth is not available).

In~\cite{Sluimer:2015}, the problem of lung CT segmentation in a supervised fashion is investigated, where a feature vector is assigned to each voxel. The vectors, containing features such as position, intensity, and derivations in $X$ and $Y$ directions, are fed to a $K$-nearest neighborhood (KNN) classifier to classify the voxels. As the name suggests, this classifier looks at the $K$-nearest neighbors of each voxel, and selects the label based on the class of the neighbors. Lung tumor segmentation in an unsupervised fashion is studied by Afshar \textit{et al.} in~\cite{Afshar:2016}, where they first segmented the lungs from the background, based on an active contour model. Consequently, the tumors were segmented using a Gustafson-Kessel clustering approach~\cite{Lv:2004}, which takes the pixels' feature vectors as inputs. This clustering method has a key advantage over the $K$-means clustering, in that it does not produce clusters with the same shapes.

In summary, conventional lung tumor segmentation methods suffer from the aforementioned drawbacks limiting their generalization capability for lung tumor segmentation. In particular, although utilization of pre-designed features might be suitable for distinguishing the tumor from the boundary in special cases, it is not generalizable to different types, locations, and sizes of the tumors. Next, we review deep learning-based methods that relax the need for constructing and defining pre-designed features.

\subsection{Deep learning-based segmentation methods} \label{sec:deep}
\begin{table*}[t!]
\caption{\small Brief description of the proposed methods in the $\LOTUS$ benchmark.}
\label{table:brief}
\centering
\begin{tabular}{ | l | p{7cm} |  p{7cm}|}
\hline
~~~~~~\textbf{Team}&~~~~~~~~~~~~~~~~~\textbf{Description of the Method}& ~~~~~~~~~~~~~~~~~~~\textbf{False Positive Reduction}\\
\hline
Markovians & A ``'LungNet'' architecture, containing convolutional layers and dilated convolutions to downsample the input. Output of all the layers are concatenated, along with the input before going through the dense layer. & Regions with a below-threshold area are removed.\\
\hline
Spectrum & A ``Recurrent 3D-DenseUNet'', which is a combination of DenseNet, CNN, and RNN to capture the inter-slice dependencies. & An adaptive thresholding based on the histogram of the output mask probability.\\
\hline
NTU-MiRA& A ``'DenseHighRes3DNet'' which incorporates dilated convolutions, residual connections, and skip connections. & NA\\
\hline
BUET Kaio-ken& A U-Net which takes the original slices and their Daubechies one wavelets as inputs. & A Random Forest classifier is trained based on the statistical features extracted from the output of the U-Net.\\
\hline
PolyUTS& A U-Net is trained with a dice loss to improve the segmentation accuracy.& A Gradient Boosting classifier is trained based on the features learned by the U-Net.\\
\hline
IITH& A ``Deeplabv3'' incorporating dilated convolutions and spatial pyramid pooling. & Partial Least Squares Discriminant Analysis on the original slices to detect tumor-containing ones before going through the ``Deeplabv3''.\\
\hline
\end{tabular}
\end{table*}
Deep learning-based lung tumor segmentation techniques are mainly developed based on Convolutional Neural Networks (CNNs), which are deep architectures resulted from stacking convolutional, pooling, and fully connected layers. The CNNs have shown promising results in different domains and applications such as medical image segmentation. The success of these networks arises from their ability to extract hierarchical discriminative features, without prior knowledge on the types of features. Furthermore, the convolutional layers use shared weights across the input, which reduces their computational cost. To solve the problem of tumor segmentation~\cite{Ciompi:2017}, CNNs can be adopted in two main fashions~\cite{Messay:2010}. The first approach is to treat the segmentation problem as a classification one, and classify pixels (i.e., a two class problem) through the fully connected layers of the CNN. The second trend is to develop a Fully Convolutional Network (FCN), where there are no fully connected layers, and the up-sampling layers are responsible to maintain the original image size. U-Net~\cite{Ronneberger:2015} is a type of the FCN, and has shown promising results in biomedical image segmentation. This architecture is an encoder-decoder network, where there are several skip pathways between the layers. These skip connections have the ability to maintain the details of the objects and lead to a fine-grained segmentation.

Although deep learning-based segmentation has been shown to be successful in medical imaging applications, and tumor segmentation in particular, its applicability is limited by the fact that these networks require relatively huge amounts of data, which is typically difficult to obtain in the medical domain. Furthermore, for the problem of tumor segmentation, the training dataset needs to be carefully annotated by the experts, which is time-consuming and not always feasible. One solution to this shortcoming is to augment the dataset~\cite{Setio:2017}, using different transformations and rotations. Two other commonly used solutions to the problem of segmentation in the absence of large amount of data are semi-supervised and weakly-supervised training~\cite{Feng:2017}. In a semi-supervised case, there are both annotated and non-annotated images in the dataset, where the unlabeled portion is used to learn the distribution of the data in a generative manner. On the other hand, in the weakly supervised case, only image-level labels (whether a tumor is present in the image) are available, and pixel-level labels can be accessible through learning the image regions that are most important for discriminating the image-level labels. Transfer learning~\cite{Shin:2016} is another widely used approach, where, typically, a natural image dataset is used to train the network, and the medical images are, consequently, utilized to fine-tune the network.
Finally, there are also hybrid approaches to tumor segmentation, where different algorithms within the conventional and/or deep learning-based segmentation are combined, with the goal of increasing the accuracy and robustness of the results. For instance, reference~\cite{Mukherjee:2017} has combined deep learning and the traditional graph-cut method to handle the complicated shapes of the lung nodules.

\section{Image Segmentation Benchmarks}\label{sec:LTB}
Challenges are, typically, aimed to provide a platform with reference datasets and robust evaluation metrics, where different algorithms can be evaluated and compared. In other words, during a challenge, all groups utilize the same training dataset to optimize their processing algorithms, and the same test dataset is used to measure the associated performance. This kind of approach is in contrast with the situation, where researchers train and test their proposed algorithm on the datasets of their choice, impeding a fair judgment between their solutions. Capitalizing  on the advantages of benchmarks, especially in medical areas, where accurate solutions are of significant importance, there have been a number of challenges seeking solid and meticulous methods to solve different problems in medical areas. In the following section, a couple of competitions on the problem of lung tumor segmentation are specified, along with their differences with the current $\LOTUS$ benchmark.

\subsection{Previous Competitions Related to Lung Tumor/Nodule}\label{sec:Pre}
``Data Science Bowl 2017\footnote{\url{https://www.kaggle.com/c/data-science-bowl-2017#description}}'' is a relatively recent challenge, focusing on lung cancer detection. However, unlike the $\VIP$ , the problem description of this competition was limited to determining the presence of tumor in the CT images, and tumor segmentation was beyond its scope.

Another key challenge designed for the problem of lung tumor analysis was the ``Lung Nodule Analysis 2016 (Luna16\footnote{\url{https://luna16.grand-challenge.org/Home/}})'', whose goal was to evaluate the performance of several state-of-the-art algorithms in lung tumor detection and false positive reduction, using the LIDC-IDRI dataset, available through the Cancer Imaging Archive~\cite{Armato:2011,Clark:2013}. In the detection phase, different possible nodule locations are identified, whereas in the false positive reduction phase, all candidates are investigated to remove the non-nodule ones. The $\VIP$ 2018 differs from the Luna16 in the following two major ways:
\begin{enumerate}
\item Luna16 dealt with the problem of lung nodule detection, in screening CT scans, where the goal was to identify the location of the nodule, rather than its fine-grained boundary, which is the objective of the LOTUS benchmark.
\item The performance of the algorithms proposed in Luna16 was assessed based on a Receiver Operating Characteristic (ROC) curve, to measure the sensitivity and specificity of the methods. However, the metrics used in $\VIP$ were based on measuring how much the predicted boundary agrees with the actual boundary from the ground truth. In other words, the objective of the $\VIP$ challenge was to measure the accuracy of the segmentation, rather than predicting only the presence of the tumor in a detected location.
\item The $\VIP$ was restricted to undergraduate students supervised by graduate students and faculty members.
\end{enumerate}

Another challenge dealing with the problem of lung tumor segmentation is called ``Automatic Nodule Detection 2009 (ANODE09)''~\cite{Ginneken:2010}. This competition was different from the $\VIP$, in the sense that it evaluated the segmentation accuracy based on their distance from the center of an actual nodule. In other words, unlike the $\VIP$, which evaluated the exact boundaries of the tumors, ANODE09 considered the rough estimation of the tumor position.

This completes a brief overview of other challenges similar to the $\VIP$. Next, we present the setup of the $\LOTUS$ benchmark.

\section{Setup of the $\LOTUS$ Benchmark}\label{sec:setup}
\begin{figure}[t!]
\centering
\includegraphics[width=.5\textwidth]{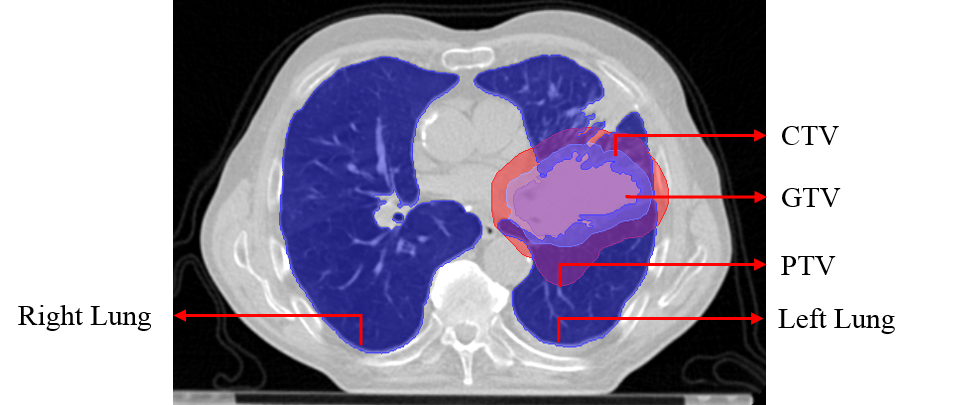}
\caption{\small Annotation of different parts of the lung CT image.}\label{fig:annot}
\end{figure}
The $\LOTUS$ benchmark was established through the 2018 $\VIP$ competition, which was organized in conjunction with 2018 IEEE International Conference on Image Processing (ICIP). In this section, we present the competition setup, problem description, data preparation and annotation, evaluation criteria, and finally the number of participants along with a brief introduction to the proposed methodologies. The challenge description is also available through its web-page: \url{http://i-sip.encs.concordia.ca/datasets.html}.

During the spring 2018, participating teams had to register for the $\VIP$ through the competition webpage, to have access to the dataset, described later in Section~\ref{sec:data}. Teams were considered eligible if they were consisted of: (i) One faculty member to supervise the team; (ii) At most one graduate student as the mentor, and; (iii) At least three, but no more than ten undergraduate students. From a total of $28$ registered teams, from several countries, $6$ teams (introduced in the Appendix) successfully completed all components of the competition, among which, based on the evaluation criteria, explained later in Section~\ref{sec:eval}, $3$ teams made it to the final competition stage. The final stage was held on October 2017, in Athens, Greece, during the ICIP 2018 conference.  Table~\ref{table:brief} outlines the methods proposed by the $6$ final participating teams. These methods are explained in more detail in Section~\ref{sec:detail}. As it can be inferred from Table~\ref{table:brief}, all the proposed solutions include a deep learning-based technique as the major segmentation step. The results of these proposed techniques are presented in Section~\ref{sec:EXP}. Next, the problem description is introduced.

\subsection{Problem description}\label{sec:data}
As stated previously, the $\LOTUS$ benchmark focuses on the problem of lung tumor segmentation from CT images. The first challenge is that not all image slices associated with the CT scan of a patient contains a tumor. Furthermore, there may be several regions in an image, having the potential to mistakenly be recognized as tumors (false positives). Successful methods, therefore, are the ones that not only provide accurate segmentation, based on the criteria defined in Section~\ref{sec:eval}, but also involve fewer false positives.

\subsection{LOTUS dataset}\label{sec:data}
\begin{figure}[t!]
\centering
\includegraphics[width=.4\textwidth]{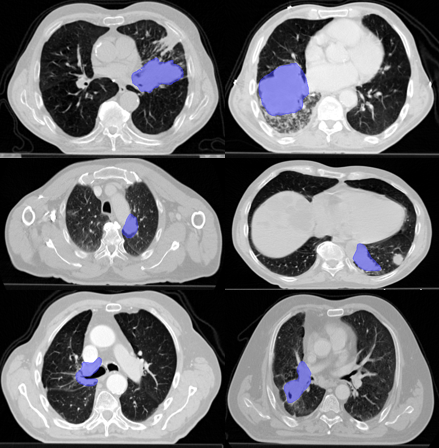}
\caption{\small Examples of the annotation of the gross tumor region.}\label{fig:example}
\end{figure}
\begin{table*}[t!]
\centering
\caption{\small Total number of CT scans and 2D plane slices.}
\label{tab:machine}
\begin{tabular}{c c|c|c|c|c|}
\cline{3-6}
 & &\multicolumn{2}{|c|}{\textbf{CT scanner}}
 & \multicolumn{2}{|c|}{\textbf{Number of slices}}\\
\hline
 \multicolumn{1}{|c|}{Dataset} & Total number of patients  & CMS Imaging Inc. & Siemens  & Tumor  & Non tumor \\
\hline
\multicolumn{1}{|c|}{Training} & $260$ & $60$ & $200$ & $4296 (13.7\%)$ & $26951 (86.3\%)$  \\
\hline
\multicolumn{1}{|c|}{Validation} & $40$ & $34$ & $6$ & $848(18.9\%)$ & $3610 (81.1\%)$  \\
\hline
\multicolumn{1}{|c|}{Test} & $40$ & $0$ & $40$ & \multicolumn{2}{|c|}{$5115$}  \\
\hline
\end{tabular}
\end{table*}
\newcolumntype{C}[1]{>{\centering}p{#1}}
\begin{figure*}
\centering
\begin{forest}
for tree={
  if level=0{align=center}{
    align={@{}C{45mm}@{}},
  },
  grow=east,
  draw,
  edge path={
    \noexpand\path [draw, \forestoption{edge}] (!u.parent anchor) -- +(5mm,0) |- (.child anchor)\forestoption{edge label};
  },
  parent anchor=east,
  child anchor=west,
  l sep=5mm,
  tier/.wrap pgfmath arg={tier #1}{level()},
  edge={ultra thick, rounded corners=1pt},
  fill=white,
  rounded corners=1pt,
  drop shadow,
}
[Total: $100$
[Validation set: $30$
[Dice score:$10$]
[Mean-surface distance: $10$]
[$95\%$ Hausdorff distance: $10$]
]
[Test set: $60$
[All slices: $30$
[Dice score :$10$]
[Mean-surface distance: $10$]
[$95\%$ Hausdorff distance: $10$]
]
[Tumor-containing  slices: $30$
[Dice score: $10$]
[Mean-surface distance: $10$]
[$95\%$ Hausdorff distance: $10$]
]
]
[Report: $10$]
]
\end{forest}
\caption{The scoring paradigm of the LOTUS competition.}\label{fig:para}
\end{figure*}
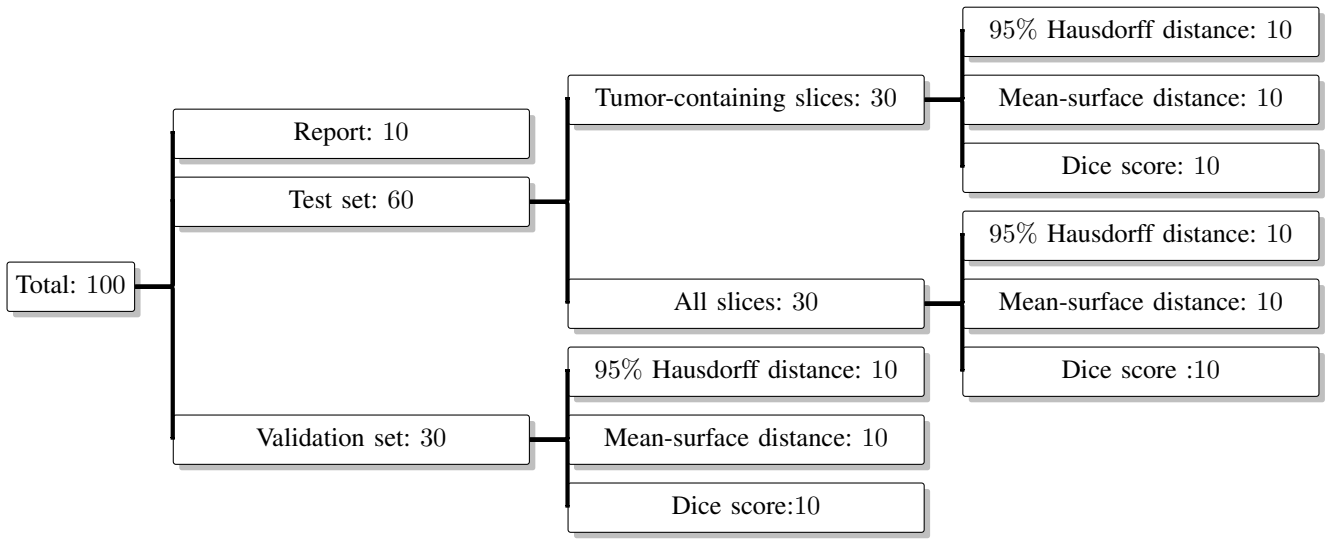

The introduced dataset during the $\VIP$ competition is an updated and modified version of the NSCLC-Radiomics black{dataset}~\cite{Aerts:2014,NSCLC,TCIA}, retrieved from the Cancer Imaging Archive. The new dataset, also provided by our collaborators at the MAASTRO Clinic (Maastricht, the Netherlands), includes the following two major improvements:
\begin{enumerate}
\item[1.] First, the initial dataset (NSCLC-Radiomics) contained several cases ($108$ patients to be exact) that were not annotated. Besides, in some cases the annotations were misplaced. These flaws are fixed in the new dataset.
\item[3.] Second, the new dataset includes the annotation of not only the main tumor area, but also several other organs, such as the left and right lung, and different tumor volumes including the gross tumor volume (GTV), the clinical target volume (CTV), and the planning tumor volume (PTV), as shown in Fig.~\ref{fig:annot}.
\end{enumerate}
The $\LOTUS$ benchmark involves the segmentation of the GTV, examples of which are shown in Fig.~\ref{fig:example}. The described dataset has been made available to the participants in three separate sets as outlined below:
\begin{itemize}

\item \textbf{\textit{Training Set}}: This set contains both the CT images and annotations for $260$ subjects (patients), with the goal of being used for training the segmentation models.
\item \textbf{\textit{Validation Set}}: Similar to the training set, the validation set includes both the images and annotations. However, this set is smaller (containing $40$ subjects), and the participants had to use it for the validation of their models. The result of the validation (reported by the teams) is incorporated in the final evaluation, as described next in Section~\ref{sec:eval}.
\item \textbf{\textit{Test Set}}: The CT images of $40$ subjects are allocated to test the proposed methods. Participants did not have access to the annotations for this set. Testing of the proposed models is performed by the organizers, and the results are incorporated for the final evaluation.
\end{itemize}
Table~\ref{tab:machine} further shows the number of CT scan slices in each of the aforementioned three datasets. Next, we present the evaluation criteria incorporated in the $\LOTUS$ benchmark.
\subsection{Evaluation Metrics}\label{sec:eval}
The evaluation was performed based on both the reported performance on the validation set and the calculated performance on the test dataset, which was carried out by comparing the binary segmentation masks, provided by participants, with the ground truth. The following three evaluation metrics were adopted for measuring the performance in both cases (on validation set by participants, and on the test set by organizers):

\begin{enumerate}
\item \textit{Dice Score}: This score measures the relative overlap of the ground truth and the predicted region, taking value $1$ as its maximum, when the two regions have a complete agreement, and minimum value of $0$, when there is no overlap between the two underlying regions. The dice score is defined as follows
\begin{eqnarray}\label{eq:c1}
\text{Dice score} = \frac{2(|\X|\cap|\Y|)}{|\X|+|\Y|},
\end{eqnarray}
where $\X$ and $\Y$ correspond to the set of voxels belonging to the ground truth and the predicted region, respectively. The symbol $\cap$ represents the intersection operation, and $|\cdot|$ denotes the size of the set. In the situations where there is no tumor in a slice, but the proposed model outputs a non-zero segmentation mask (i.e., a false positive), value of zero is assigned to the dice score. The final score is the average over all slices and subjects.
\item \textit{Undirected Average Hausdorff Distance}: This measure (mean-surface distance) is the average of the two directed Hausdorff distances  $\vec{d}_{H,\mathrm{avg}}(\X,\Y)$ and $\vec{d}_{H,\mathrm{avg}}(\Y,\X)$, defined as follows
\begin{eqnarray}\label{eq:c2}
d_{H,\mathrm{avg}}(\X,\Y) = \frac{\vec{d}_{H,\mathrm{avg}}(\X,\Y) + \vec{d}_{H,\mathrm{avg}}(\Y,\X)}{2}.
\end{eqnarray}
The directed Hausdorff distance $\vec{d}_{H,\mathrm{avg}}(\X,\Y)$ calculates the distance between two boundaries $\X$ and $\Y$, by averaging over the distance between points in $x \in \X$ and their closest points in $y \in Y$, as defined below
\begin{eqnarray}
\vec{d}_{H,\mathrm{avg}}(\X,\Y) = \frac{1}{|\X|} \sum_{x \in \X} \min_{y \in \Y} d (x,y),
\end{eqnarray}
where $d (x,y)$ is the distance between two points $x$ and $y$. To be consistent with the dice score, the inverse of the mean-surface distance is considered, and $0$ is assigned to the false positives.
\item \textit{The $95\%$ Hausdorff Distance}: This measure is calculated similar to the mean-surface distance, with the difference that instead of averaging over the distance between points in $\X$ and their closest points in $\Y$ to calculate the $\vec{d}_{H,\mathrm{avg}}(\X,\Y)$, the point in $\X$ is selected, whose distance to its closest point in $\Y$ is greater than or equal to exactly $95\%$ of the other points in $\X$.
\end{enumerate}
\noindent
\textit{\textbf{Ranking Procedure}}: Ranking of the teams participated in the 2018 $\VIP$ is performed based on the three described criteria (Dice score, mean-surface distance, and $95\%$ Hausdorff distance). Furthermore, a relatively small score is considered for the reports teams wrote, summarizing their algorithm and results, to encourage participants to follow a well-organized research process. Fig.~\ref{fig:para} shows the final scoring paradigm used to rank the teams. For each metric adopted within the validation or the test set, score of $5-10$ is assigned to teams, with $10$ representing the best result and $5$ denoting the worst. For the validation set, only the tumor-containing  slices are considered (there is no penalty for false positives), whereas for the test set, two cases are taken into account. In the first case, all slices are considered, meaning that false positives decrease the score. In the second case, only tumor-containing  slices are considered. In the first round of the $\LOTUS$ benchmark, three teams were selected, based on the described process, to compete in the second round, where for the same test dataset, teams were provided with the indices of the tumor-containing slices. In other words, the focus of the second round was segmentation of the tumor rather than detecting the slices with tumors. Therefore, the three finalists had the opportunity to improve their segmentation models, knowing which slices contained the tumors.

\subsection{Limitations of the LOTUS Benchmark and Lessons Learned}
\begin{figure}[t!]
\centering
\includegraphics[width=0.2\textwidth]{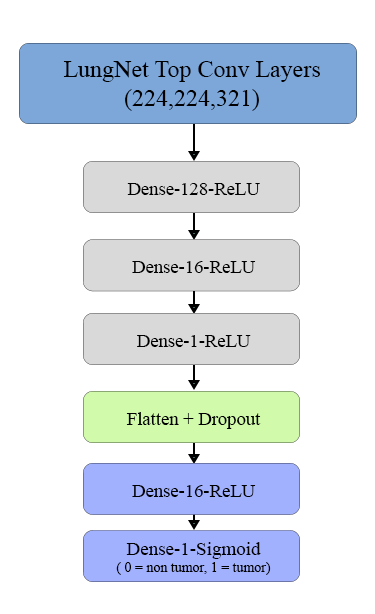}
\caption{\small The classification network of the team Markovian. Figure from team Markovian with permission. \label{fig:mar1}}
\vspace{-.2in}
\end{figure}
\begin{figure}[t!]
\centering
\includegraphics[width=0.5\textwidth]{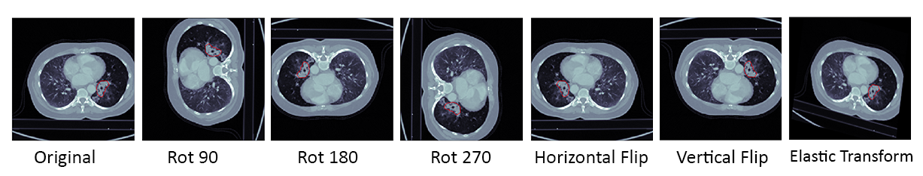}
\caption{\small Seven-fold augmentation including rotation (Rot) with 90,180 \& 270 degrees, horizontal and vertical flip, and elastic transformation. \label{fig:Aug}}
\vspace{-.2in}
\end{figure}
The LOTUS benchmark has provided researchers with the opportunity to train and test their lung tumor segmentation models on a unique dataset, and based on pre-determined metrics. There are, however, some weaknesses that need to be considered in future competitions. One of such shortcomings is that the evaluation criteria are based on the distance and overlap of the ground truth and predicted region. These metrics do not take the shape of the regions into account, meaning that they do not care about the similarity between the two underlying shapes. Second, although the introduced dataset is a completed and corrected version of the NSCLC-Radiomics dataset, it still needs to be expanded to be counted as a reliable and representative set of different shapes and locations of tumors. Furthermore, since the tumor region has, typically, a fuzzy boundary, annotations need to be provided by several experts. Last but not least, the introduced dataset comes from a couple of different sources and scanners, leading to a few inconsistencies between the images. For the ease of use and interpretation, images need to be standardized.

\section{Proposed $\VIP$ Segmentation Algorithms over $\LOTUS$ } \label{sec:detail}

\begin{figure}[t!]
\centering
\includegraphics[width=0.3\textwidth]{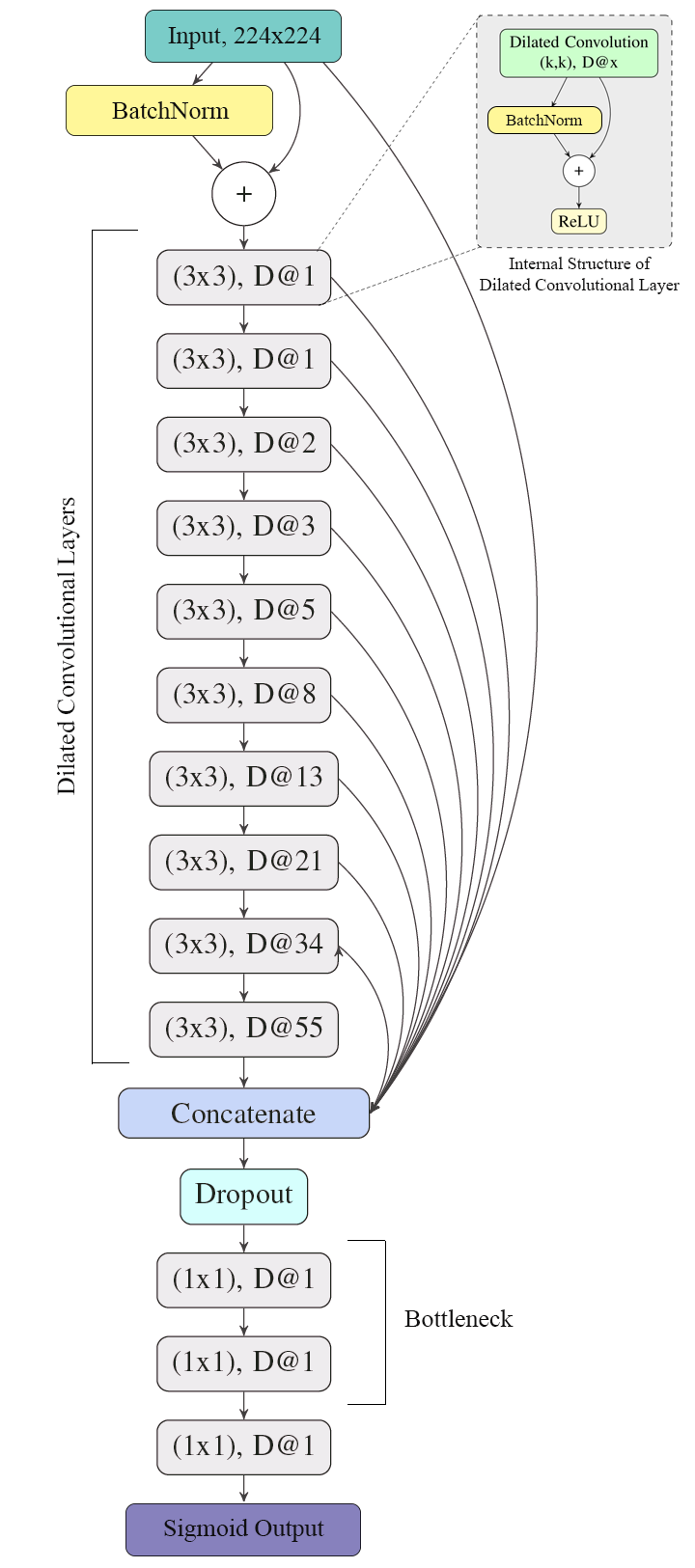}
\caption{\small The segmentation network of the team Markovian. Figure from team Markovian with permission. \label{fig:mar2}}
\vspace{-.1in}
\end{figure}

\begin{figure*}[t]
\centering
\mbox{\subfigure[]{\includegraphics[width=0.33\textwidth]{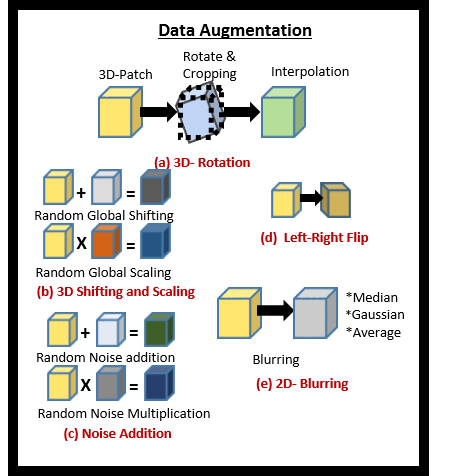}}
\subfigure[]{\includegraphics[width=0.7\textwidth]{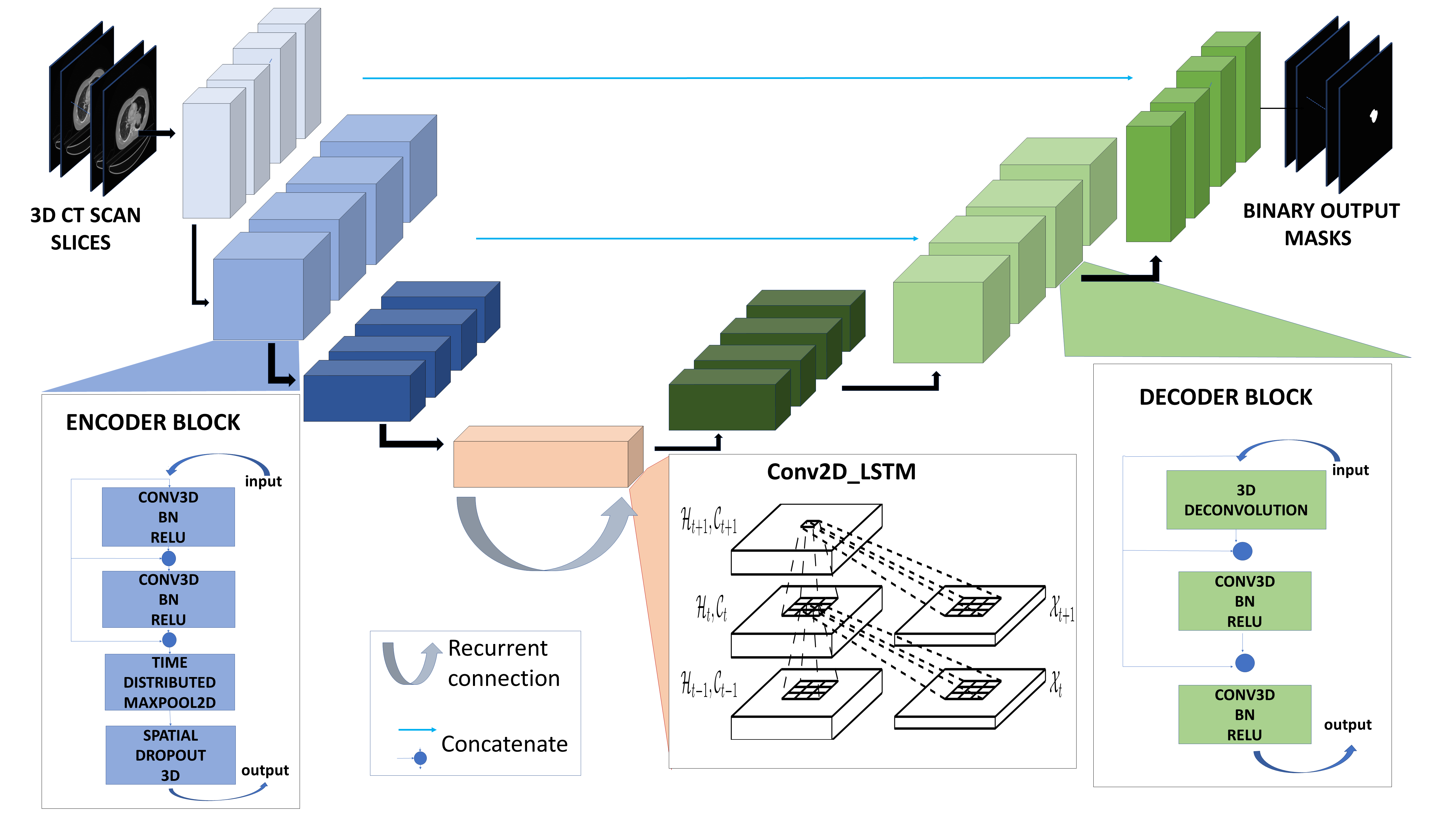}}}
\caption{\small (a) Data augmentations incorporated by team Spectrum. (b) The segmentation network of the team Spectrum. Figure from team Spectrum with permission. \label{fig:spe1}}
\vspace{-.1in}
\end{figure*}

\begin{figure}[t!]
\centering
\includegraphics[width=0.5\textwidth]{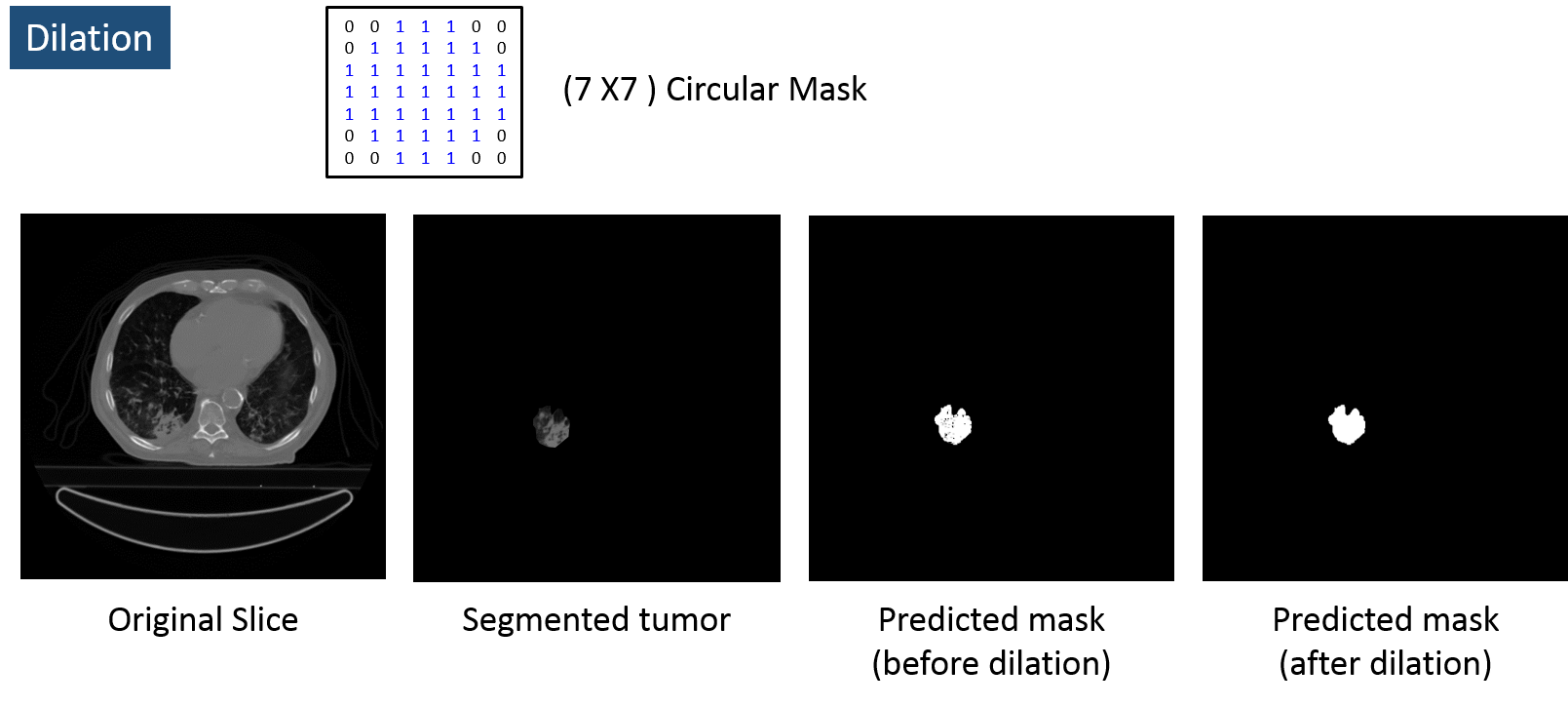}
\caption{\small Post processing with dilation proposed by team Spectrum. \label{fig:Dilation}}
\vspace{-.2in}
\end{figure}

In this section, the methods proposed by the teams participated in the IEEE VIP-CUP 2018 are briefly described.
\subsection{Team Markovians}
The method proposed by team Markovians consists of the following four main blocks~\cite{Hossain:2019}:
\begin{enumerate}
\item[1.] Pre-processing including data normalization and augmentation.
\item[2.] Binary classification to distinguish between slices containing tumor and those without any abnormality.
\item[3.] Segmentation, which is the main block of the proposed approach.
\item[4.] Post-processing to enhance the quality of the segmentation.
\end{enumerate}
In the following sub-sections, we present each of the aforementioned four steps in more details.
\subsubsection{\textbf{Pre-processing}}
The main purpose of the pre-processing step is to make the images consistent and standard. Since the CT images are obtained from two different scanners, the pixels lie within different intensity ranges. Therefore, intensities are normalized before going through the rest of the steps. Furthermore, after analyzing the training data, all images are cropped to remove empty spaces that do not correspond to a body organ.  Second phase of the pre-processing step considers the problem of over-fitting, which is caused by the lack of large enough dataset (common in almost all medical domains). As shown in Fig.~\ref{fig:Aug}, to overcome this problem, training images are augmented through several operations, such as rotation, flipping, and transformation. Lastly, to address the class imbalance problem caused by the fact that the number of images not containing a tumor is significantly bigger than the number of images containing a tumor, the negative instances are sub-sampled.

\subsubsection{\textbf{Binary Classification}}
For classifying images as tumor-containing  or not, the same network that is trained in the segmentation phase is adopted. This network will be explained in Section~\ref{sec:segMar}. To adjust this network, which is trained with the aim of producing the segmented image, to the classification problem at hand, five fully-connected layers, as shown in Fig.~\ref{fig:mar1}, are added to the network, where the final layer is a classification one with Sigmoid activations. The resulted network is fine-tuned for classification and distinguishing between tumor-containing and not-tumor-containing images.

\subsubsection{\textbf{Segmentation}}\label{sec:segMar}
The deep network architecture for the segmentation of the lung tumors, shown in Fig.~\ref{fig:mar2}, is a ``LungNet'' adopted from~\cite{Anthimopoulos:2018}. This network has a fully convolutional structure and the downsampling layers are replaced with dilated kernels, which increases the receptive fields. The output of the first $10$ layers and the input are all concatenated before going through the dropout and the rest of the network. This concatenation can serve the network in two ways: (i) It makes an easier flow of gradient, and; (ii) It allows the integration of features from various details and abstractions. The original LungNet architecture~\cite{Anthimopoulos:2018} incorporates instance normalization after each convolutional layer, where instance normalization refers to normalizing each individual instance in the batch, independent from the other instances. However, in the architecture proposed by team Markovian this type of normalization is replaced with batch normalization, where normalization factors are computed based on all of the instances within the mini-batch. Team Markovian has used the negative log of the dice-score (Eq.~\eqref{eq:c1}) as the network loss function to be minimized through back propagation.

\subsubsection{\textbf{Post-processing}}
In the post-processing step, the goal is to remove noisy pixels and reduce the false positive cases. Two fundamental morphological operations, i.e., dilation and erosion, are applied to the output of the network.
Here, dilation refers to expanding the shapes in an image, using a structuring element. In other words, dilation adds to the pixels of the boundaries. In contrast, the erosion operation removes pixels on the boundaries.
Consequently, contours are detected and the area of the regions within the contours are calculated, and regions that are associated with a below-threshold area, are removed as false positives.

\subsection{Team Spectrum}
The segmentation pipeline developed by team Spectrum can also be described in four main steps of pre-processing, segmentation, false positive reduction, and post-processing. In the pre-processing step, images are normalized and augmented (as shown in Fig.~\ref{fig:spe1}(a)) through applying rotation, shifting, noise addition, flipping, and blurring. In the post-processing step, as shown in Fig.~\ref{fig:Dilation}, the generated masks go through a dilation morphological operation with a circular kernel (structuring element) to improve the intersection over union (IOU).
Below the other two main phases, i.e., segmentation and false positive reduction, are explained.
\subsubsection{\textbf{Segmentation}}\label{sec:segspe}

\begin{figure*}[t]
\centering
\includegraphics[width=.98\textwidth]{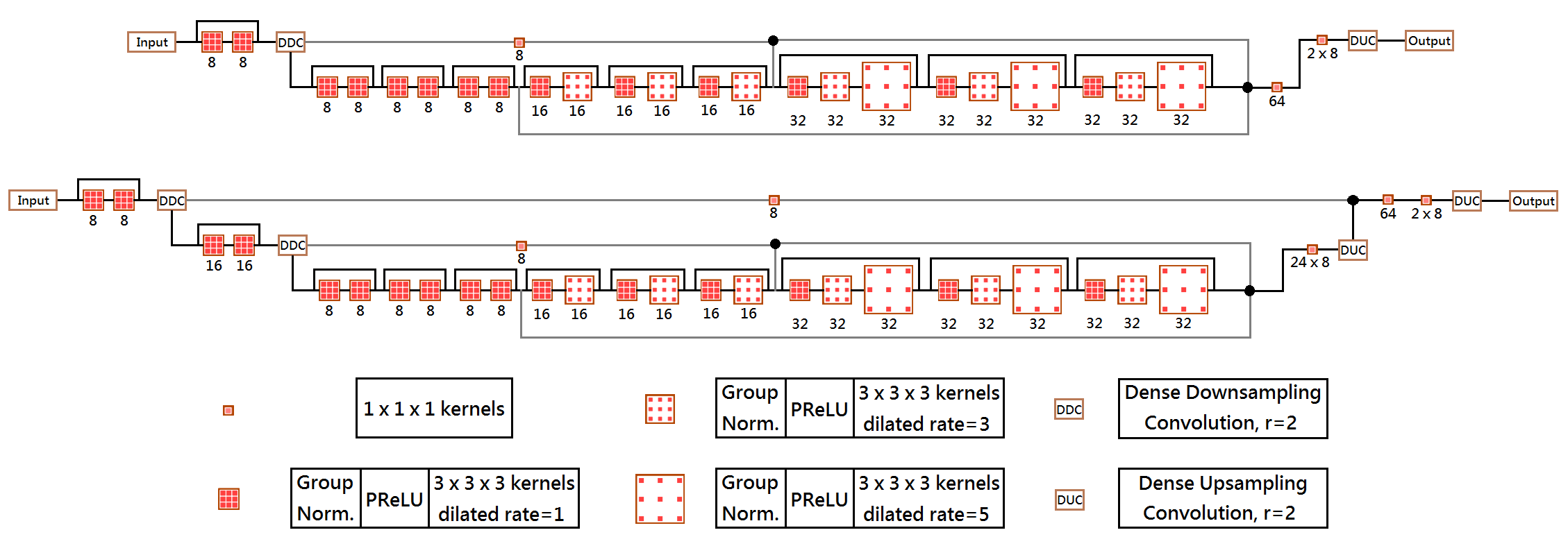}
\caption{\small The segmentation network of the team NTU-MiRA. Figure from team NTU-MiRA with permission. \label{fig:ntu1}}
\end{figure*}
A novel architecture named ``Recurrent 3D-DenseUNet'', shown in Fig.~\ref{fig:spe1}(b), is proposed for the segmentation of the lung tumors. At training time, the inputs to this network are stack of slices containing tumors, whereas in the test time, inputs are all slices, including tumor-containing and not-tumor-containing ones. Therefore, outputs of the network, at the test time, go through a false positive reduction phase, which will be described in~\ref{sec:spefa}.

 The proposed network is a combination of three fundamental networks: DenseNet~\cite{Huang:2017}, UNet, and convolutional recurrent network. In the Recurrent 3D-DenseUNet, the 3D pooling layers are replaced with 2D ones, to avoid the loss of information along channel axis, i.e., lung image slices. The most important property of this network is that 3D inputs (stack of tumor slices) are utilized, to capture inter-slice dependencies. Therefore, to accurately incorporate these dependencies, convolutional long-short-term-memory (LSTM) blocks are used in the transition from the encoder to the decoder network. Since, these blocks are placed after the encoder part of the network, they have the capability to capture the dependencies between the high-level features. Furthermore, as the dimension of the data is decreased after going through several convolutional and pooling layers, training of the LSTM blocks is more computationally efficient. Finally, $(1-\text{dice-score)}$ is used as the loss function of the whole network, to be minimized through back propagation.

\subsubsection{\textbf{False Positive Reduction}}\label{sec:spefa}
An adaptive dynamic thresholding approach is presented for removing the regions that are incorrectly predicted as belonging to a tumor. This threshold is applied to the output probabilities of the segmentation network, described in section~\ref{sec:segspe}. To build the adaptive threshold, the histogram of pixel intensities of the 3D probability output of the segmentation network is generated, and the intensities corresponding to the $3^{th}$, $4^{th}$, $5^{th}$, and $6^{th}$ highest peaks of the histogram are averaged with empirical weights.  The relation, below, shows the mathematical formulation of the dynamic threshold $I_d$
\begin{equation}
I_d=\frac{1}{N-2}\Big[\sum_{i=3}^N a_i \times arg I(h_i)\Big]\quad \forall h_i \in H,
\end{equation}
where $N$ is the index for the last ordered histogram peak, $H$ is the histogram sorted in a descending order, $argI(h_i)$ represents corresponding intensity at the peak $h_i$, and finally $a_i$ are weighting coefficients.

\subsection{Team NTU-MiRA}
\begin{table*}[t!]
\centering
\caption{\small Results of the proposed methods by the $6$ participating teams in the first stage.}
\label{tab:res1}
\begin{tabular}{c|c|c|c|c|c|c|}
\cline{2-7}
 & \multicolumn{3}{|c|}{\textbf{Validation Set}}
 & \multicolumn{3}{|c|}{\textbf{Test set: All Slices}}\\
\hline
 \multicolumn{1}{|c|}{Participating} & Dice  & Mean-surface & $95\%$ Hausdorff  & Dice  & Mean-surface & $95\%$ Hausdorff \\
 \multicolumn{1}{|c|}{teams} & score  &  inverse &  inverse  & score  &  inverse &  inverse  \\
\hline
\multicolumn{1}{|c|}{Markovian} & $0.626$ & $0.098$ & $0.053$ & $0.02$ & $0.007$  & $0.004$\\
\hline
\multicolumn{1}{|c|}{Spectrum} & $0.74$ & $0.55$ & $0.13$ & $0.032$ & $0.009$ & $0.005$ \\
\hline
\multicolumn{1}{|c|}{NTU-MiRA} & $0.60$ & $0.13$ & $0.064$ & $0.032$ & $0.009$ & $0.004$ \\
\hline
\multicolumn{1}{|c|}{BUET Kaio-ken} & $0.55$ & $0.0009$ & $0.0009$ & $0.053$ & $0.008$ & $0.004$ \\
\hline
\multicolumn{1}{|c|}{PolyUTS} & $0.44$ & $0.1$ & $0.04$ & $0.024$ & $0.007$ & $0.003$\\
\hline
\multicolumn{1}{|c|}{IITH} & $0.68$ & $3.12$ & $0.46$ & $0.01$ & $0.004$ & $0.002$ \\
\hline
\end{tabular}
\end{table*}
The lung tumor segmentation algorithm developed by the team NTU-MiRA consists of the two phases of pre-processing and segmentation, where in the pre-processing step, images are augmented through random cropping, elastic deformation, and spatial scaling. Furthermore, to handle the class imbalance problem, training is performed in two serial rounds. In the first round the proposed segmentation network, which will be discussed in Section~\ref{sec:NTU:D}, is trained with the objective of minimizing the dice score, whereas in the second round, the loss function is changed to a focal loss defined as
\begin{eqnarray}
q_i &=& \begin{cases}
p_i, & \text{if $y_i=1$}\\
1-p_i, & \text{otherwise}\end{cases}\\
\text{Focal Loss} &=& -\alpha(1-q_i)^{\gamma} log(q_i),
\end{eqnarray}
where $y_i$ is the ground truth, $p_i$ is the output of the model, and $\alpha$ and $\gamma$ are hyper parameters.

\subsubsection{\textbf{Segmentation}}\label{sec:segntu}
The segmentation model proposed by team NTU-MiRA (referred to as the Dense-HighRes-3DNet), shown in Fig.~\ref{fig:ntu1}, is based on the model presented in~\cite{Li:2017}, which has two important properties, i.e.,
\begin{itemize}
\item Dilated convolution and residual connections are incorporated. Residual connection, referring to the merging of the output of a block with its input, have several advantages. For instance, these connections contribute to a more effective training by smoothing the information propagation.
\item The receptive fields of the kernels are reduced to generate less distorted borders in the segmentation task.
\end{itemize}
The difference between the proposed Dense-HighRes-3DNet and the model presented in~\cite{Li:2017} is that, similar to an encoder-decoder network, the former incorporates downsmapling and upsampling layers, before and after the main residual network, respectively. Furthermore, dilated convolutions are replaced with hybrid dilated convolutions (HDC)~\cite{Wang:2018}, where a range of dilation rates are applied and concatenated to overcome the gridding issue in a dilated convolution. In advance, Dense-HighRes-3DNet includes not only several residual connections, but also skip connections to further prevent the information loss throughout the network.

The upsampling process, in most of the deep networks, involves either a bilinear upsampling, which is not learnable, or a deconvolution, which needs zero padding before the convolution operation. Therefore, team NTU-MiRA has proposed to use Dense Upsampling Convolution (DUC)~\cite{Wang:2018} in the decoder part of the networks. Inspired from the DUC, this team has also proposed the Dense Downsampling Convolution (DDC) for the encoder part.

\subsection{Team BUET Kaio-ken} \label{sec:NTU:D}
The method proposed by the team BUET Kaio-ken follows the pipeline of: pre-processing, segmentation, false positive reduction, and post-processing. In the pre-processing step, a heat-map of the occurrence of the tumor is generated from the training data, and all images are cropped based on this heat-map to compensate for the small number of tumorous pixels compared to the non-tumorous ones. Furthermore, to extract textural information from the images, a Daubechies one wavelet (db1) up to two level simplifications is used. Image augmentation through flipping and rotating the slices along with tumor enlarging, using morphological dilation, are also adopted in the pre-processing step. In advance, to consider the dependencies between consecutive slices, each slice is investigated along with its two immediate neighbors. Therefore, the segmentation network, which is a U-Net, takes the following images as inputs: current slice, previous slice, next slice, db1 wavelet level 1, and db2 wavelet level 2. The U-Net is trained based on a five-fold cross validation, and the final pixel probabilities are calculated based on a simple bagging of the five trained models.

In the false positive reduction phase, several statistical features, such as mean, standard deviation, sum, maximum value, skewness, and kurtosis, are computed from the output probability map of the U-Net, and a Random Forest classifier is trained to determine tumor-containing and not-tumor-containing slices. Finally, in the post-processing step, a Hysteresis thresholding is applied to enhance the quality of the output.

\subsection{Team PolyUTS}
Team PolyUTS has also adopted a U-Net for the lung tumor segmentation. To overcome the class imbalance problem caused by the relatively small size of the tumor compared to the background, \textit{cross entropy loss} is replaced with the dice loss. Rotation, flipping, and random cropping are applied to the input images, in the augmentation phase. Since the U-Net, alone, is prone to producing false positives, this team has used the deep features learned by the U-Net to train a Gradient Boosting classifier. These deep features are extracted from the feature maps of the $4^{th}$ up-convolutional layer of the U-Net. Team PolyUTS has further reduced the FPR by taking the correlation between consecutive slices into account.

\subsection{Team IITH}
The segmentation algorithm proposed by the team IITH starts with a pre-processing step, where images are normalized and lungs are segmented using a K-means clustering. Consequently, in a detection phase, the pre-processed images are analyzed to determine the tumor-containing slices. This step consists of a dimensionality reduction and a classification. Team IITH has adopted the Partial Least Squares Discriminant Analysis (PLS-DA) for this step. The main segmentation model, which is applied to the tumor-containing slices identified in the previous step, is called a ``Deeplabv3''~\cite{Chen:2017}. The Deeplabv3 has the following two important properties:
\begin{itemize}
\item First, it incorporates dilated convolutions to extract denser feature maps.
\item Second, it incorporates spatial pyramid pooling to capture objects at multiple scales, by using parallel dilated convolutions of various rates.
\end{itemize}
This completes description of the proposed and implemented processing pipelines by participants of the $\VIP$. Next, we present the results.

\section{Results} \label{sec:EXP}
\begin{table}[t!]
\centering
\caption{Results of the proposed methods by the $3$ finalist teams in the final stage.}
\label{tab:res2}
\begin{tabular}{|c|c|c|c|}
\hline
\textbf{Finalist} & \textbf{Dice-score} & \textbf{Mean-surface}  & \textbf{$95\%$ Hausdorff} \\
\textbf{Teams} &   & \textbf{Inverse} &  \textbf{Inverse}\\
\hline
Markovians & $0.59$  & $0.17$ & $0.08$ \\
\hline
NTU-MiRA & $0.54$ & $0.17$ & $0.08$\\
\hline
Spectrum & $0.52$ & $0.14$& $0.07$\\
\hline
\end{tabular}
\end{table}
Table~\ref{tab:res1} shows the results obtained by the $6$ participating teams. Results on the validation set is reported by the teams on tumor-containing slices. Results on the test set is calculated by the organizers on both the tumor-containing slices and the all-slices cases, where in the latter, false positives have negative impact on the Dice score, Mean-surface distance, and $95\%$ Hausdorff distance. Based on this table and the scoring process described in Section~\ref{sec:eval}, teams Markovians, Spectrum, and NTU-MiRA made it to the final stage.

\begin{figure}[t!]
\centering
\mbox{\subfigure[]{\includegraphics[scale=0.15]{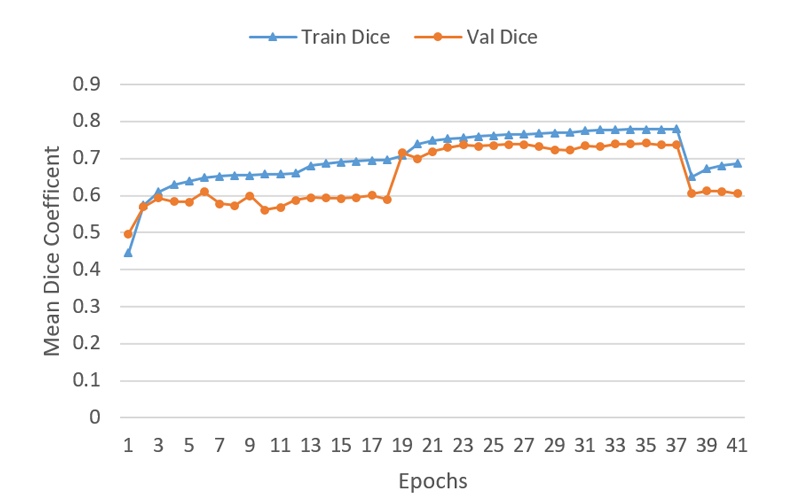}}
\subfigure[]{\includegraphics[scale=0.15]{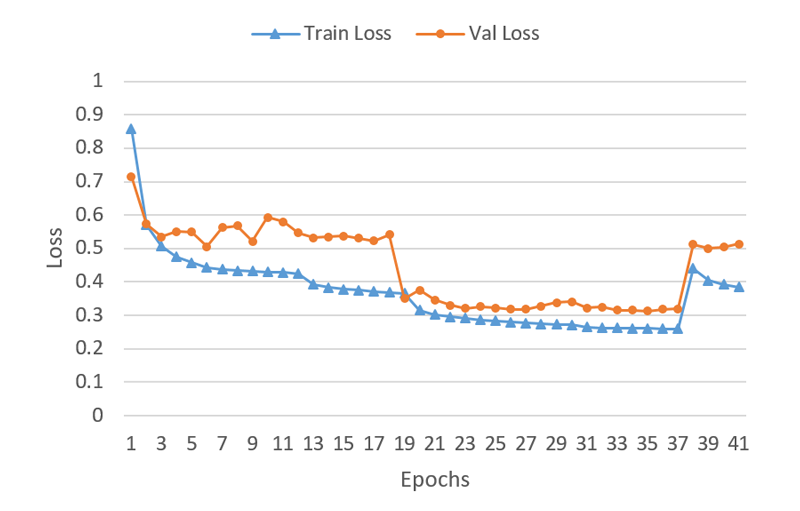}}}
\caption{\small Training phase for the segmentation model developed by Team Markovian (first ranked team). (a) Increased dice score through the training. (b) Decreased loss through the training.\label{fig:MAr-Train}}
\vspace{-.1in}
\end{figure}

\begin{figure}[t!]
\centering
\includegraphics[width=.45\textwidth]{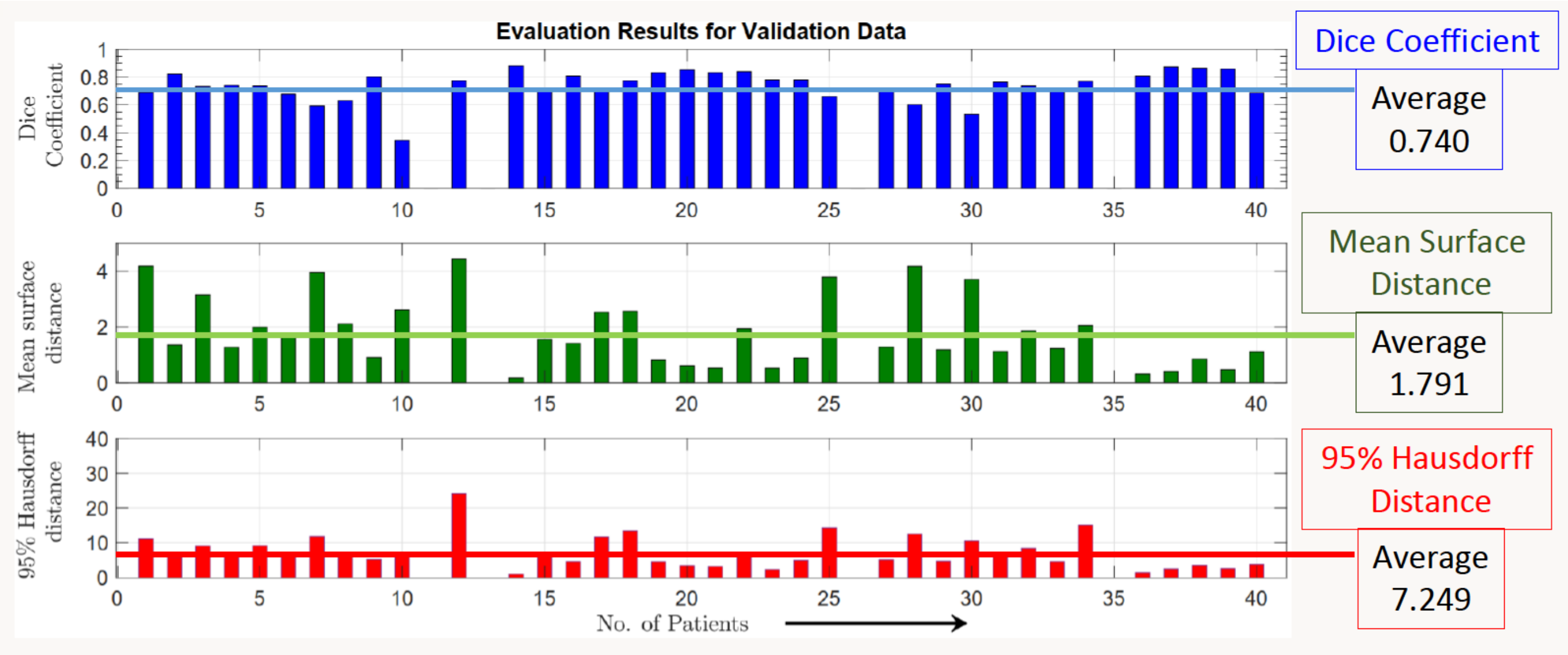}
\caption{\small Evaluation results via Team Spectrum's proposed solution. \label{fig:resSpec}}
\vspace{-.2in}
\end{figure}

Fig.~\ref{fig:MAr-Train} illustrates training phase of the segmentation model developed by the champion of the $\VIP$. The following two subsets of data are used for training purposes: (i) \textit{Subset A}, which only consists of tumor-containing-slices with augmentation, and; (ii) \textit{Subset B}, which consists of all tumor-containing-slices, ten randomly selected non-tumor slices from each patient, along with seven fold augmentation.
During the training phase, initially the segmentation model is trained based on Subset A till the validation loss plateaued (epochs 1 to 18). The model is retrained with Subset B with learning rate of 0.0001 (epochs 19 to 37).
After the loss plateaued, Subset A was again used till loss started to rise (epochs 38 to 40). It is worth mentioning that the increased dice coefficient for Subset B is due to the addition of blank masks which result in a dice coefficient of 1 if no mask is generated. Fig.~\ref{fig:resSpec} shows the results obtained based on the proposed solution by Team Spectrum.

\begin{figure}[t!]
\centering
\includegraphics[width=.45\textwidth]{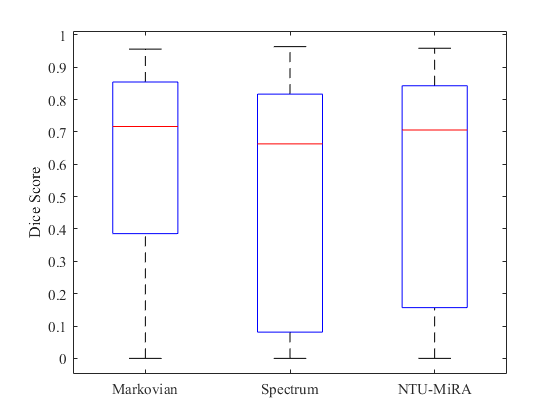}
\caption{\small The Box-plot of the obtained results for the tumor-containing slices. \label{fig:box}}
\vspace{-.2in}
\end{figure}

In the final stage, the three finalists were provided with the indices of the tumor-containing slices in the test set, and the focus was on the correct segmentation of the tumors. Table~\ref{tab:res2} and Fig.~\ref{fig:box} show the obtained results, according to which, the segmentations are satisfactory when the model is trained and tested on only the tumor-containing slices, whereas in the previous case (model trained and tested on all slices), the accuracy decreases due to the high number of false positives. In other words, more effort should be made to improve the accuracy of the tumor detection.

To test whether the obtained results have a statistically significant difference, we performed a Wilcoxon-rank-sum test, with the null hypothesis of the results having equal medians, which led to the rejection of the null hypothesis for each two combination of the teams, in a significance level of $0.05$. Finally, based on the results, reports, and the fact that the team NTU-MiRA did not have a false positive reduction phase, Markovian, Spectrum, and NTU-MiRA won the first, second, and third places respectively.

\begin{figure}[t!]
\centering
\mbox{\subfigure[]{\includegraphics[scale=0.3]{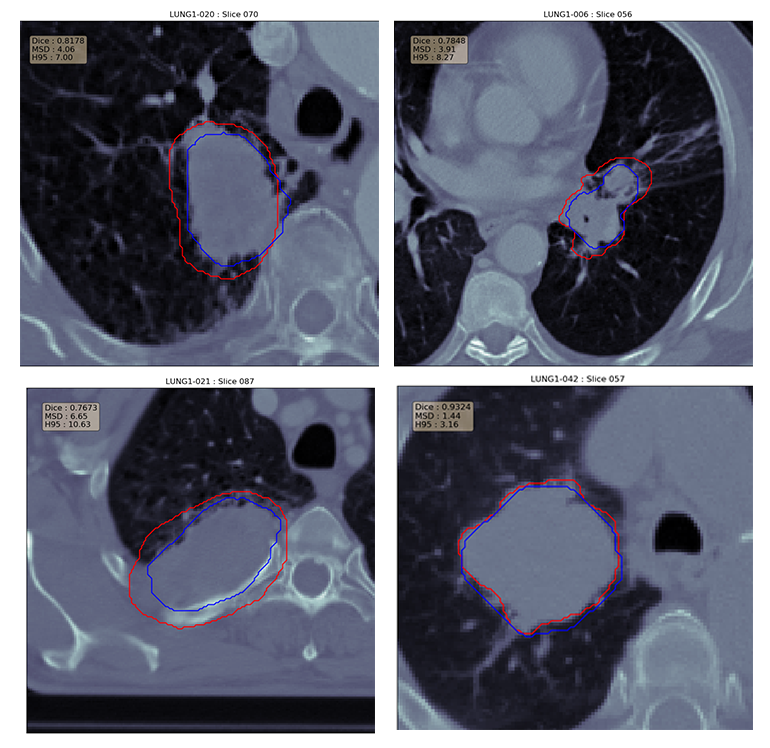}}}\\
\mbox{\subfigure[]{\includegraphics[scale=0.3]{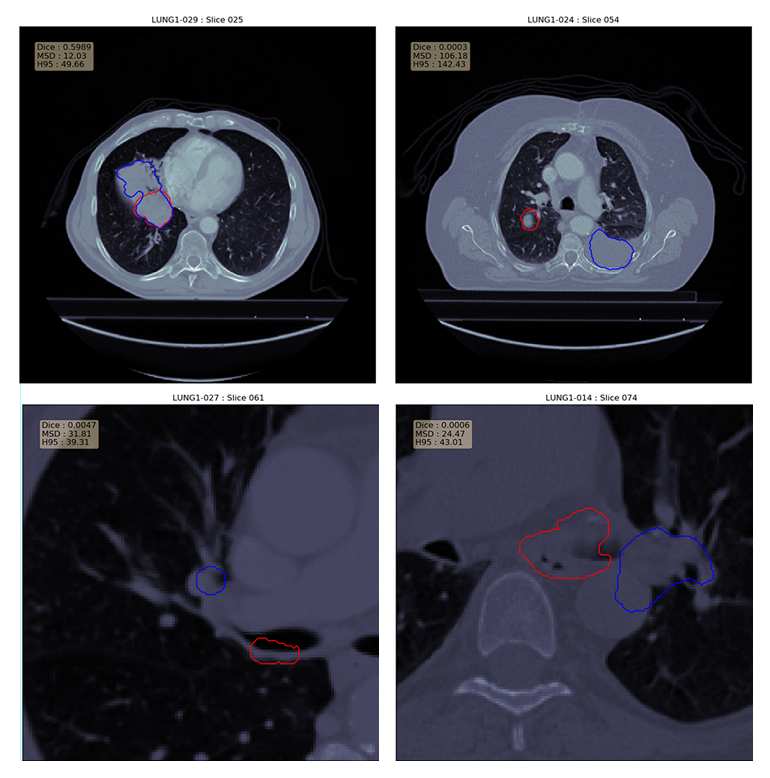}}}
\caption{\small (a) Examples of the cases, where the segmentation method is successful. (b) Examples of the cases, where the segmentation method fails. \label{fig:case}}
\end{figure}

To investigate the cases, where the segmentation method developed by the team Markovian (the winner team) succeed or fail to provide an accurate delineation, some of the segmentation results are shown in Fig.~\ref{fig:case}(a) and Fig.~\ref{fig:case}(b), along with the ground truth. As it can be inferred from these figures, in some cases the predicted boundary can perfectly encompass the tumor region, and it is even, visually, more satisfactory compared to the ground truth (Fig.~\ref{fig:case}(a)). There are also some failure cases that although the predicted boundary seems to be related to a tumor-like region, according to the ground truth, it is, in fact, a false positive (Fig.~\ref{fig:case}(b)), needing more reliable detection techniques to be excluded.

\section{Conclusion}  \label{sec:con}
In this paper, we have reported the design, implementation, and results of the LOTUS benchmark. This competition is organized to provide the researchers with the opportunity to develop and test their proposed approaches for lung tumor segmentation on a unique dataset and based on pre-defined metrics, which are missing in most of the studies in this domain. The LOTUS dataset is an updated version of the NSCLC- Radiomics dataset, where incorrect annotations are corrected, and missing annotations are provided. Based on the obtained results from the algorithms developed by participants, which are mostly the variants of different deep learning architectures, the tumor segmentation can be achieved with a satisfactory accuracy (dice score between $0.52$ and $0.59$ on the slices containing the tumor). More effort, however, should be put in detecting tumor-containing  slices and decreasing the false positives.

\section{Acknowledgment} \label{sec:ack}

This work was partially supported by the Natural Sciences and Engineering Research Council (NSERC) of Canada through the NSERC Discovery Grant RGPIN-2016-04988. This project has also been funded in whole or in part with Federal funds
from the U.S. National Cancer Institute, National Institutes of Health, under contract HHSN261200800001E. The content of this publication does not necessarily reflect the views or policies of the U.S. Department of Health and Human Services, nor does its mention of trade names, commercial products, or organizations imply endorsement by the U.S. Government.

P. Afshar, A. Mohammadi, and X. Wu are with the Concordia Institute for Information Systems Engineering, Concordia University, Montreal, QC, Canada.

K. N. Plataniotis is with the  Department of Electrical and Computer Engineering, University of Toronto, Toronto, Canada.

K. Farahani is with the Division of Cancer Treatment and Diagnosis, National Cancer Institute (NCI), Rockville, MD, USA.

J. Kirby is with the Frederick National Laboratory for Cancer Research, Frederick, MD, USA.

A. Oikonomou is with the Department of Medical Imaging, Sunnybrook Health Sciences Centre, University of Toronto, Toronto,~Canada.

A. Asif is with the Electrical and Computer Engineering Department,  Concordia University, Montreal,~Canada.

 L. Wee and A. Dekker are with the Department of Radiation Oncology, MAASTRO Clinic, Maastricht, The Netherlands; School for Oncology and Developmental Biology (GROW), Maastricht University, Maastricht, The Netherlands.

 M. A. Haque, Sh. Hossain,  M. K. Hasan. and U. Kamal are with the Department of Electrical and Electronic Engineering, Bangladesh University of Engineering and Technology, Dhaka, Bangladesh.

W. Hsu and J. Lin are with the  Department of Computer Science and Information Engineering, National Taiwan University, Taipei, Taiwan.

M. S. Rahman, N. Ibtehaz, and Sh. M. A. Foisol are with the Department of Computer Science and Engineering, Bangladesh University of Engineering and Technology, Dhaka, Bangladesh.

K. M. Lam, Zh. Guang, and R. Zhang are with the Department of Electronic and Information Engineering, The Hong Kong Polytechnic University, Hung Hom, Kowloon, Hong Kong.

S. S. Channappayya, Sh. Gupta, and Ch. Dev are with the  Department of Electrical Engineering, Indian Institute of Technology (IIT Hyderabad), India.

\bibliographystyle{IEEEtran}

%
\section{Appendix}
Table~\ref{table:par} presents the $6$ teams participated in the LOTUS competition.

\begin{table*}[t!]
\caption{Participating teams in the LOTUS competition.}
\label{table:par}
\centering
\begin{tabular}{|c|c|p{10cm}|}
\hline
Team&Supervisor&~~~~~~~~~~~~~~~~~~~~~~~~~~~~~~~~~~~~~~~~~~~~Team members\\
\hline

 \Vcentre{~~\includegraphics[width=0.07\textwidth]{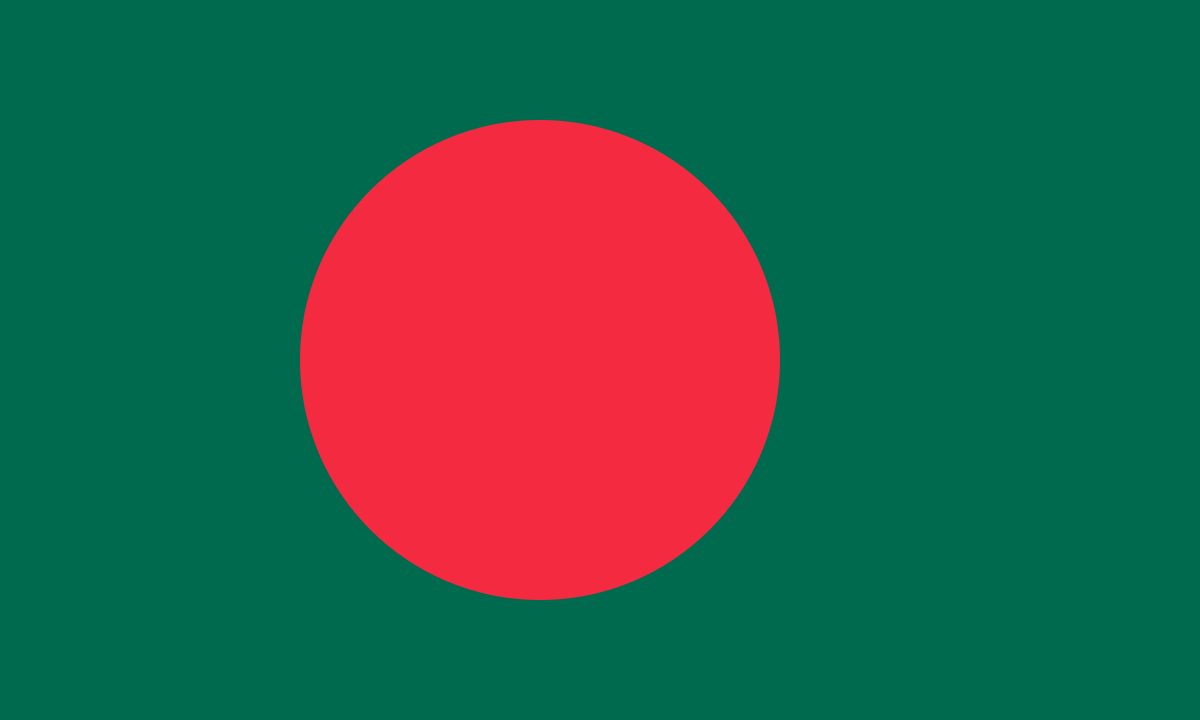}} \Vcentre{ Markovian} &  \Vcentre{Mohammad Ariful Haque}&  Shahruk Hossain, Muhammad Suhail Najeeb, Zaowad Rahabin Abdullah, Asif Shahriyar,
Md. Mushfiqur Rahman, Md. Tariqul Islam, Fahim Hafiz, Md. Monayem Hassan, Md. Farhan Shadiq, and A. K. M. Naziul Haqu\\
\hline
 \Vcentre{\includegraphics[width=0.07\textwidth]{Bangladesh.png}} \Vcentre{ Spectrum} &  \Vcentre{Md. Kamrul Hasan}&  Uday Kamal, Abdul Muntakim Raﬁ,  and Rakibul Hoque \\
 &&
\\
\hline
 \Vcentre{~~~\includegraphics[width=0.07\textwidth]{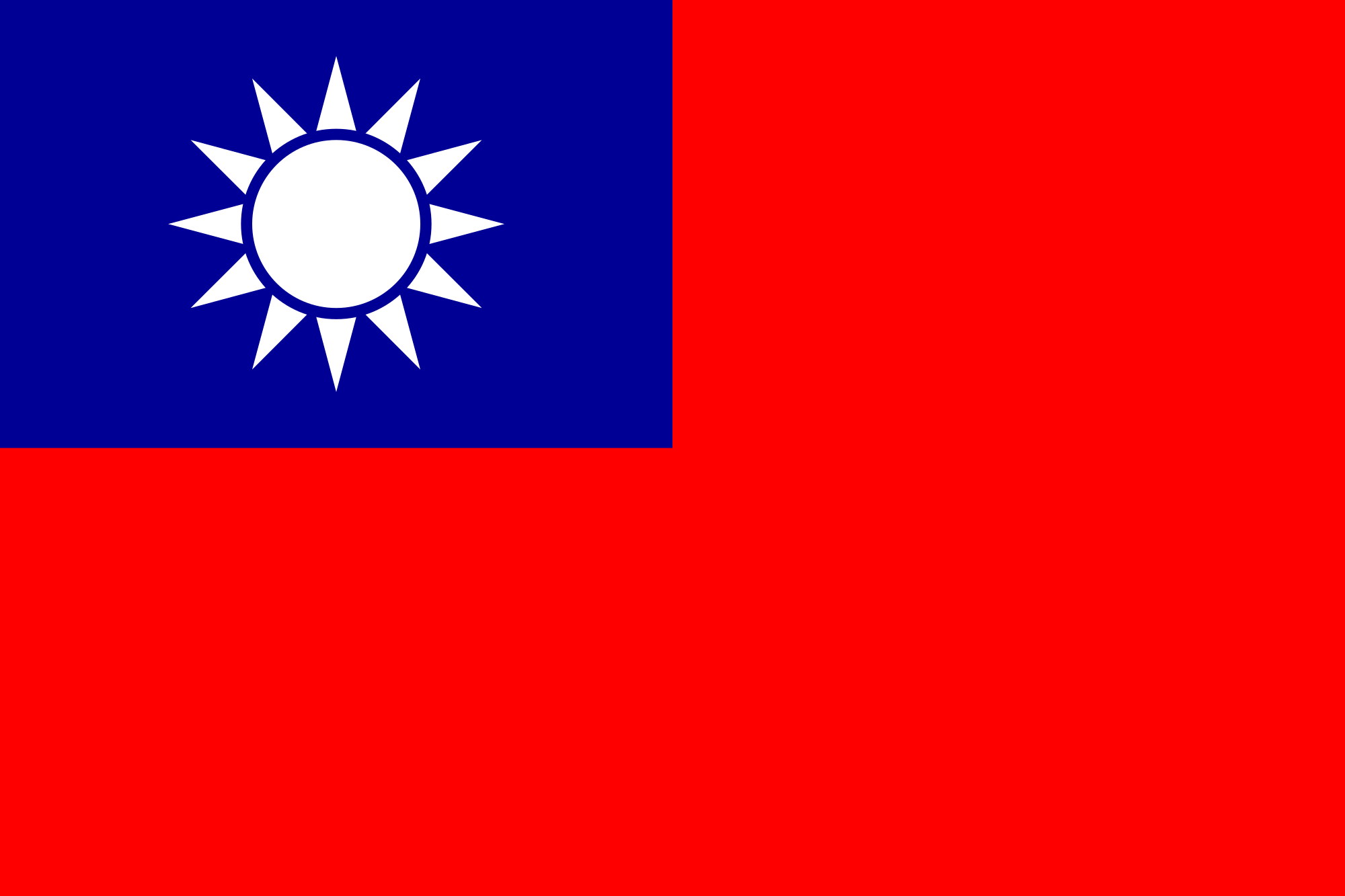}}  \Vcentre{ NTU-MiRA} &  \Vcentre{Winston Hsu}&  Jhih-Yuan Lin, Min-Sheng Wu, Yu-Cheng Chang, Yun-Chun Chen, Chao-Te Chou and Chun-Ting Wu\\
 &&
\\
\hline
 \Vcentre{~~~~~~~~\includegraphics[width=0.07\textwidth]{Bangladesh.png}}  \Vcentre{ BUET Kaio-ken} &  \Vcentre{M Sohel Rahman}& Sheikh Mostofa Amir Foisol, Debolina Halder, Trishna Chakraborty and Nabil Ibtehaz\\
 &&
\\
\hline
 \Vcentre{\includegraphics[width=0.07\textwidth]{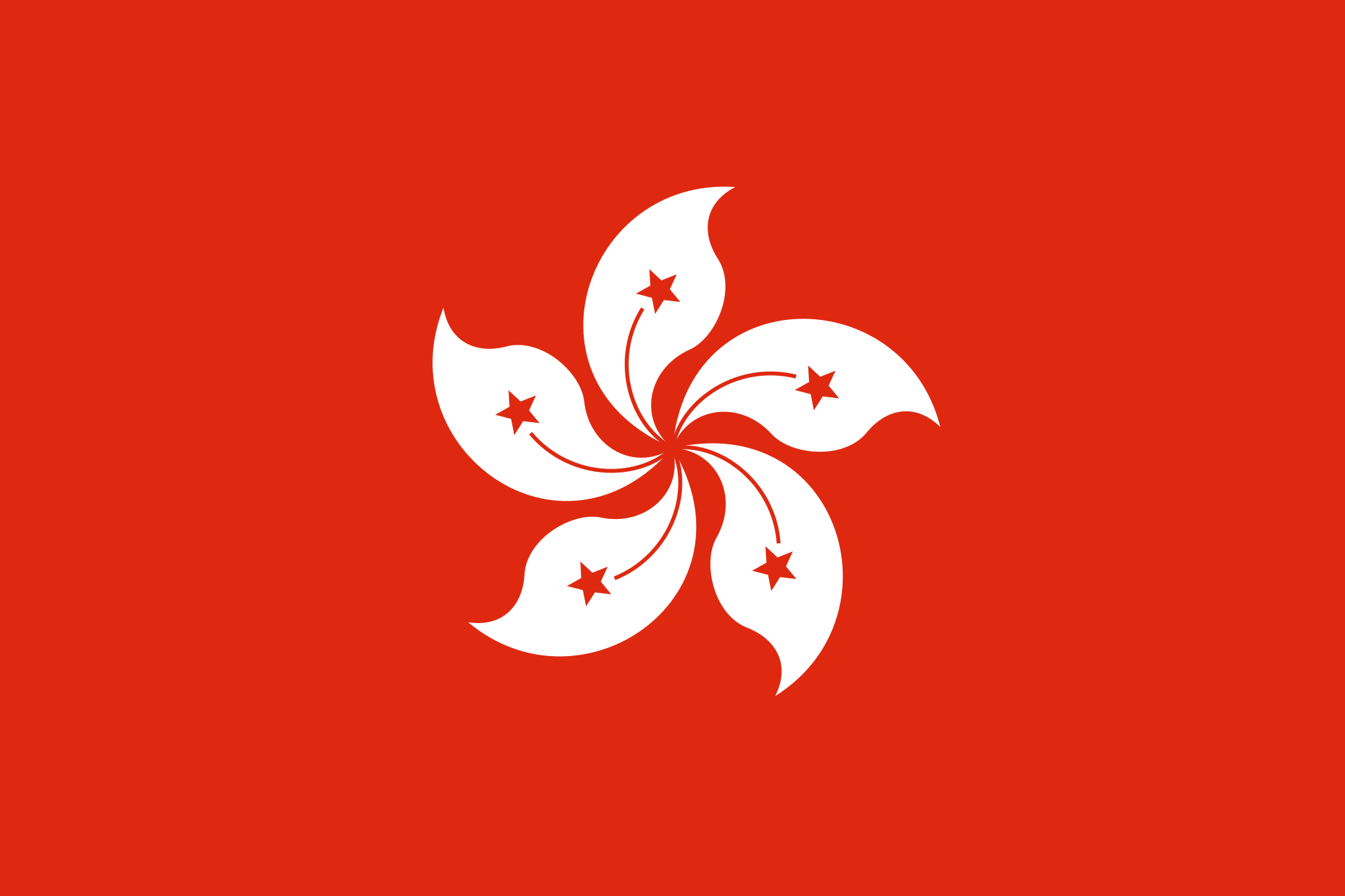}}  \Vcentre{ PolyUTS} &  \Vcentre{Kin-Man Lam}& Zhong Guan, Ye Huang, Wenqi Jia, Blake Rainey, Qingqu Xu, Chaoran Zhang, Runze Zhang and Lei Liu\\
\Vcentre{\includegraphics[width=0.07\textwidth]{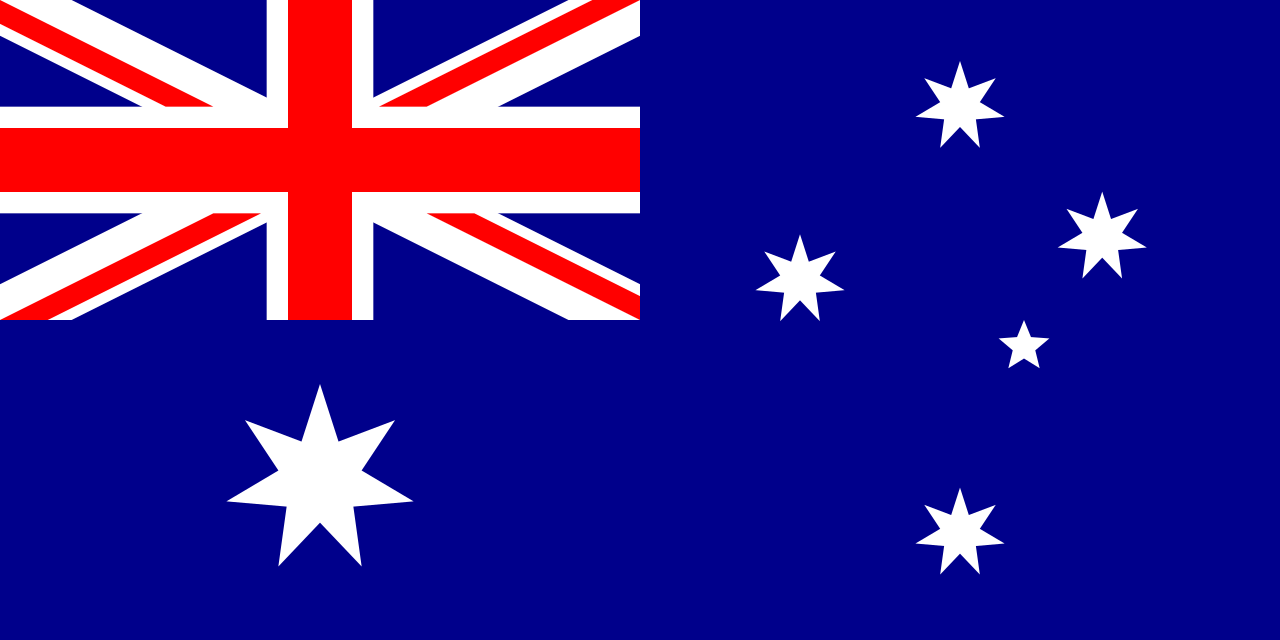}~~~~~~~~~~~~} &&
\\
&&
\\
\hline
 \Vcentre{\includegraphics[width=0.07\textwidth]{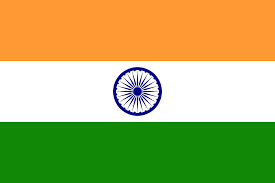}}  \Vcentre{ IITH~~~~~} &  \Vcentre{Sumohana S. Channappayya}& Chander Dev, Chakradhar Nakka, Aravind Ganesh Pathapati, Shashank Gupta, Snehith Aithu, Narayan Kothari, Havish Potaraju, Sreeja Pilli, Thanuja Battini, Vamshi Teja Racha and Charantej Reddy Pochimireddy\\
 &&
\\
\hline

\end{tabular}
\end{table*}

\end{document}